\documentclass[aps,prl,twocolumn,groupedaddress,preprintnumbers]{revtex4}
\usepackage{graphicx}
\usepackage{amsmath,amssymb,color}
\usepackage[usenames,dvipsnames]{xcolor}
\usepackage{multirow} 
\usepackage{hyperref}
\usepackage{setspace}

\def\sideremark#1{\ifvmode\leavevmode\fi\vadjust{\vbox to0pt{\vss% the remark
 \hbox to 0pt{\hskip\hsize\hskip1em%                          will appear only
 \vbox{\hsize3cm\tiny\raggedright\pretolerance10000%          on the side
 \noindent #1\hfill}\hss}\vbox to8pt{\vfil}\vss}}}%
\begin{document}
\preprint{CALT 68-2937, BRX-TH{ $\mathbf {666}$}}

%\doublespacing
\title{No consistent bimetric gravity?}

\author{S.~Deser}
\email[]{deser@brandeis.edu}
%\homepage[]{Your web page}
%\thanks{}
%\altaffiliation{}
\affiliation{Lauritsen Lab, Caltech, Pasadena CA 91125 and Physics Department, Brandeis University, Waltham, MA 02454, USA}

\author{M.~Sandora}
\email[]{sandora@ms.physics.ucdavis.edu}
%\homepage[]{Your web page}
%\thanks{}
%\altaffiliation{}
\affiliation{Department of Physics, University of California, Davis, CA 95616, USA}

\author{A.~Waldron}
\email[]{wally@math.ucdavis.edu}
%\homepage[]{Your web page}
%\thanks{}
%\altaffiliation{}
\affiliation{Department of Mathematics, University of California, Davis, CA 95616, USA}

\begin{abstract}
We discuss the prospects for a consistent, nonlinear, partially massless (PM), gauge symmetry of bimetric gravity (BMG).
Just as for  single metric massive gravity, we show that consistency of BMG relies on it having a PM extension; we then argue that it cannot.\end{abstract}

\maketitle

\section{Introduction}

Two-tensor, bimetric (or ``$f$-$g$'') gravity~(BMG) was born as the fusion of the strong-interaction resonances of the 1960's (specifically that of a massive spin 2 field) with  General Relativity's~(GR) massless metric~\cite{Salam}. The idea was to join the two symmetrically by
giving each its own Einstein-Hilbert action, then coupling the two resulting ``spaces''
through a non-derivative ``mass term''  that necessarily reduced the
separate coordinate invariances to a single one, each field transforming as a tensor (and the mass term as a  scalar density). 
It was soon realized that upon setting, by hand, one metric to be a background, BMG could be reduced to a single met\-ric-, but non-geometric, massive gravity (mGR) theory with explicit mass terms  such that the linear limit yielded the free, Fierz--Pauli (FP), field~\cite{WZ}. 
These models enjoy the isometries of the background metric as symmetries (so are Lorentz invariant massive spin~2 theories in Minkowski backgrounds).
However, interest was soon brought to a halt
when they were found to propagate ghost modes at nonlinear level~\cite{BD}, even for generic mass terms which reduced linearly to FP (coupled to a massless graviton for BMG).
The obstruction was the appearance of a sixth---necessarily ghost---massive ``bulk'' degree of freedom~(DoF). It was only some forty years later that this obstruction 
was averted~\cite{deRham} by the discovery of---exactly~3---mass terms for which only 5  massive DoF remain, all non-ghost (one of the preferred mass terms was actually proposed in~\cite{WZ} soon after~\cite{Salam},
and had been explicitly exempted in the no-go list of~\cite{BD}). This ``no-ghost'' result was quickly extended to BMG with the same class of mass terms~\cite{Hassan}. These results opened an instant, now major industry--see~\cite{HB} for a review.

Philosophically, evasion of the classical  mGR no-go  theorem would be unsatisfactory, as it would demote  GR from its (Yang--Mills-like) status as an isolated theory---without permitted massive  neighbors---whose local symmetry cannot be broken by hand.
However, absence of a sixth massive, ghost-like mode is only a necessary, but not sufficient, condition for the ultimate consistency of mGR.
In fact, it was rapidly realized that these improved mGR models exhibited superluminality in their auxiliary sector and decoupling limit~\cite{Gruzinov,Burrage}, and quite generally suffered (local) acausalities and tachyonic modes~\cite{DW,DSW} whose interactions with (hitherto) normal matter entail catastrophic consequences\footnote{For technical reasons only, these latest no-go results leave open one class of mass terms (cubic in the background metric); they most likely suffer from the same propagating tachyons.}~\cite{Johnson,Velo,DWin}. 
Separately, detailed analysis of mGR's cosmological properties showed it to exhibit ghost instabilities about its homogeneous solutions~\cite{MU}. 
The one remedy that might have saved mGR also failed; it cannot be made partially massless (PM): Recall, in background de Sitter (dS) spaces, there exist irreps of spin 2 with only 4, rather than the 5, DoF of flat space, when the mass and~$\Lambda$ are suitably tuned. Indeed, not only is the offending mode removed in PM, but the four remaining, helicity (2,1) DoF all propagate exactly on the dS light cone, further linking PM with causal behavior~\cite{PM}.
However, the PM limit for single metric, two derivative theories is subject to various no-go results: First, a study
of PM-gauge invariant vertices gave a no-go result at quartic order~\cite{Zinoviev}. An analysis of possible truncations of Weyl-squared, conformal gravity (CG)---whose spectrum consists of relatively ghost graviton and PM modes~\cite{Maldacena,Joung}---to its PM sector also yielded a negative result~\cite{Joung}. Finally, two independent groups~\cite{DSW,deRhamPM} showed that  PM limit of
mGR does not exist. Indeed, the very (constraint) terms responsible for acausal mGR propagation~\cite{DW} 
also imply the absence of a PM Bianchi identity and hence of any accompanying  gauge symmetry~\cite{DSW}.

At this point then, mGR is relegated at best to an effective theory of dubious physical relevance. On the other
hand BMG, although at grave risk of suffering similar difficulties, might still conceivably be viable. In particular, since mGR solutions in a given background
do not solve the BMG field equations, the mGR no-go results are not directly applicable\footnote{The PM mGR no-go  does restrict possible PM gauge invariances of the higher derivative model proposed in~\cite{HassanCG} in which, 
by solving algebraically for one metric in terms of the other, BMG is  converted to a single metric model involving an 
infinite tower of higher derivative interactions. The authors of~\cite{HassanCG} claim this theory could constitute a renormalizable ghost-free model and,  even more optimistically, that a PM gauge symmetry could protect the structure of these against deadly radiative corrections. We  exhibit the pitfalls facing these claims in Section~\ref{problems}.}.
As evidence for a PM BMG model, in~\cite{HassanCG} a vacuum solution admitting a PM linearization was exhibited. 
A healthy linearization being only the very first step for constructing consistent interactions,  given the problems faced
by the PM limits of mGR and CG, one could reasonably infer  their non-existence in BMG.
On the other hand, a PM limit of BMG would likely shield it from the mGR's acausality difficulties because not only would a PM gauge invariance 
remove the offending helicity zero mode, but it would  also imply absence of the field-dependent first derivative terms (that were responsible for mGR acausalities)  in the scalar constraint
of neighboring untuned (non-PM) BMG models. 
Non-existence of a non-linear PM gauge invariance of BMG is our subject.

\section{The model}

The dynamical fields are a pair of metric tensors~$(g_{\mu\nu},\bar g_{\mu\nu})$; their action has a
cosmological Einstein-Hilbert term for each, plus a ``mass term''~$V$ depending algebraically on~$(g,\bar g)$:
\begin{widetext}
$$
S=M^2 \int \sqrt{-g}\,  \big(R(g)+2\Lambda\big) \ +\  \mu^4\! \int V(g,\bar g) \ 
+ \ \bar M^2 \int \sqrt{-\bar g}\,  \big(R(\bar g)+2\bar \Lambda\big)\, .
$$

The (mass-dimension) parameters~$(M,\bar M)$ and $\mu$ encode, respectively,  the two Planck, and the FP, masses.

The equations of motion are

\begin{equation}\label{eom}
{\cal G}_{\mu\nu}:=G_{\mu\nu}(g)-\Lambda g_{\mu\nu} -\frac{\mu^4}{M^2}\,  \tau_{\mu\nu} = 0 = G_{\mu\nu}(\bar g)-\bar\Lambda \bar g_{\mu\nu} -\frac{\mu^4}{\bar M^2} \, \bar\tau_{\mu\nu}=:\bar{\cal G}_{\mu\nu}\, ,
\end{equation}
\end{widetext}
where~$\tau_{\mu\nu}$ and~$\bar \tau_{\mu\nu}$ are, respectively (up to metric determinant factors), the ~$g$ and~$\bar g$ variations of the mass term.

The analysis of preferred mass terms is greatly simplified by working in terms of a corresponding pair of vierbeine 
$(e_\mu{}^m,f_\mu{}^m)$, rather than metrics:
$$
g_{\mu\nu}=e_\mu{}^m\eta_{mn} e_\nu{}^n\, ,\qquad \bar g_{\mu\nu}=f_\mu{}^m\eta_{mn} f_\nu{}^n\, .
$$
Spin (as well as metric) connections are the Levi-Civita (Christoffel) ones:~$\omega_\mu{}^m{}_n:=\omega_\mu{}^m{}_n(e)$ and 
$\bar\omega_\mu{}^m{}_n:=\omega_\mu{}^m{}_n(f)$.
We fix half of the  local~$SO(3,1)\times SO(3,1)$ vierbein ambiguity by requiring\begin{equation}\label{symmetry}
\varphi_{\mu\nu}:= f_{\mu}{}^m e_{\nu m}=f_{\nu}{}^m e_{\mu m}\, .
\end{equation}
These six symmetry conditions are consistent with the field equations, but leave a further six redundancies
encoded by the algebraic (St\"uckelberg) gauge symmetry 
$$
e_\mu{}^m \mapsto \Lambda^m{}_n e_\mu{}^n\, ,\qquad f_\mu{}^m \mapsto \Lambda^m{}_n f_\mu{}^n\, .
$$
(This is the diagonal subgroup of the two local Lorentz symmetries.) The underlying DoF count is 
unchanged, the~$10+10=20$ metric components have been replaced by~$16+16=32$ vierbein components
subject to~$6$ conditions~(\ref{symmetry}) and~$6$ algebraic gauge redundancies.
Before analyzing the DoF count, we specialize to the ``preferred'' mass term
\begin{equation}\label{V}
V\sim \epsilon_{mnrs}\,  e^m\wedge e^n\wedge f^r\wedge f^s\, .
\end{equation}
The other pair of allowed couplings, ~$\epsilon_{mnrs} e^m\wedge f^n\wedge f^r\wedge f^s$ and 
$\epsilon_{mnrs}  e^m\wedge e^n\wedge e^r\wedge f^s$, are not relevant for the putative PM 
limit suggested in~\cite{HassanPM}. The PM mGR studies of~\cite{DSW,deRhamPM} ruled these terms out there; this implies that they are very unlikely to be allowable PM couplings in BMG either.

For inverse vierbeine we employ~$e^\mu{}_m$ and~$f^\mu{}_m$:
$$
e_\mu{}^m e^\mu{}_n=\delta^m_n=f_\mu{}^m f^\mu{}_n\, ,
$$
while
$$(\varphi_{\mu\nu})^{-1}=e^{\mu}{}_m f^{\nu}{}_m=:\chi^{\mu\nu}=\chi^{\nu\mu}\, .
$$
In the absence of a metric
tensor, vectors and covectors can no longer be identified, while in the presence of two metrics this identification is 
no longer unique. There are various situations where we do want to use solely one or the
other metric (and its vierbeine)  to manipulate indices. 
When these arise we will write~$\stackrel{g}=$ or~$\stackrel{\bar g}=$.
For example, given a pair of covectors~$v_\mu$
and~$w_\nu$, we will write~$X\stackrel{g}= v^\mu w_\mu$ to denote the equation~$X=g^{\mu\nu} v_{\mu} w_\nu$
and~$X\stackrel{\bar g}=v^\mu w_\mu$ for~$X=\bar g^{\mu\nu} v_{\mu} w_\nu$. In mixed situations, we will indicate explicitly how 
indices are handled.

\section{The vector constraint and Bianchi identity}

For any choice of tensor density~$V(g,\bar g)$, the model has a single manifest dynamical diffeomorphism invariance, under which
\begin{eqnarray*}
\delta g_{\mu\nu}&=&\pounds_\xi g_{\mu\nu}=\nabla_\mu \xi_\nu + \nabla_\nu \xi_\mu \, ,\\[2mm]
\delta g_{\mu\nu}&=&\pounds_\xi \bar g_{\mu\nu}=\bar \nabla_\mu \xi_\nu + \bar \nabla_\nu \xi_\mu \, .
\end{eqnarray*}
(Here~$\nabla$ and~$\bar \nabla$ are  the Levi-Civita connections of~$g$ and~$\bar g$, respectively.)
This local  invariance immediately implies the vector Bianchi identity\begin{equation}\label{vBianchi}
M^2\, \sqrt{-g} \, g^{\mu\nu} \nabla_\mu {\cal G}_{\nu\rho} + \bar M^2\, \sqrt{-\bar g} \,\bar g^{\mu\nu} \bar \nabla_\mu \bar{\cal G}_{\nu\rho} = 0\, .
\end{equation}
It also reduces the (apparent) DoF count to~$10+10-2\times 4= 12$ in the usual way, but there are still further constraints.

For the particular choice of mass term~(\ref{V}), using~(\ref{symmetry}), we have
\begin{equation*}
\tau_{\mu\nu}\ \stackrel{g}=\ 2(\varphi_{\mu\rho}-g_{\mu\rho} \varphi_\sigma^\sigma)\varphi^\rho_\nu
+g_{\mu\nu}(\varphi_\rho^\rho\varphi_\sigma^{\sigma}-\varphi_\rho^{\sigma}\varphi_\sigma^\rho)
\, ,
\end{equation*}
and the same structure for~$\bar \tau_{\mu\nu}$, with all quantities barred.
Taking divergences, after some algebra (see~\cite{DSW}), we learn
\begin{widetext}
\begin{eqnarray}
\frac{M^2}{\mu^4}\, {\cal C}_\mu\ :=\ g_{\mu\nu}\chi^{\nu\rho}\nabla^\sigma {\cal G}_{\rho\sigma}
&\stackrel{g}=&-\, 2\,  \varphi^{\nu\rho}
\big(K_{\nu\rho\mu}-g_{\nu\rho} K_{\sigma}{}^\sigma{}_\mu
+g_{\mu\nu}K_{\sigma}{}^\sigma{}_\rho\big)=0\, ,\nonumber\\[2mm]
\frac{\bar M^2}{\mu^4}\, \bar{\cal C}_\mu\ :=\ \bar g_{\mu\nu}\chi^{\nu\rho}\bar \nabla^\sigma \bar {\cal G}_{\rho\sigma}
&\stackrel{\bar g}=&+\,2\, \varphi^{\nu\rho}
\big( K_{\nu\rho\mu}-\bar g_{\nu\rho}  K_{\sigma}{}^\sigma{}_\mu
+\bar g_{\mu\nu} K_{\sigma}{}^\sigma{}_\rho\big)=0\, .\label{vectors}
\end{eqnarray}
\end{widetext}
The contorsion~$K$ appearing above is  the difference of spin connections,
$$
K_\mu{}^m{}_n:=\omega_\mu{}^m{}_n(e)-\omega_\mu{}^m{}_n(f)\, .
$$
It measures the failure of parallelograms of one metric to close with respect to the other.
In particular~$\nabla_{[\mu} f_{\nu]}{}^m = K_{[\mu}{}^m{}_{|n|} f_{\nu]}{}^n$ and~$\bar \nabla_{[\mu} e_{\nu]}{}^m = -K_{[\mu}{}^m{}_{|n|} e_{\nu]}{}^n$.

Although~${\cal C}_\mu$ and~$\bar{\cal C}_\mu$ appear to furnish the theory with two separate vector constraints,
they are not independent, thanks to the vector Bianchi identity~(\ref{vBianchi}).
Thus, at this juncture, by virtue of the  single vector constraint, the DoF count sits at~$8=10+10-2\times 4-1\times 4$. To arrive (covariantly)
at the claimed generic~$7=5+2$ modes of a massive+massless ``graviton'' pair, we need to uncover exactly one, further, scalar, 
constraint. Before performing that analysis, we detour slightly to analyze the  linearization of BMG.

\section{Linearized BMG}\label{linear}

To linearize BMG, one first needs physically interesting vacua. Ref.~\cite{HassanCG} (whose linear analysis we confirm here) proposed that these can be obtained by considering tuned metrics
$$
\bar g_{\mu\nu}=\gamma \, g_{\mu\nu}\, ,
$$
where the strictly positive constant~$\gamma$ is  be fixed by the equations of motion. Then, taking~$g$ and (therefore also)~$\bar g$ to be Einstein metrics, we have 
$$
G_{\mu\nu}(g)-\lambda\,  g_{\mu\nu}=0=G_{\mu\nu}(\bar g)-\lambda\,  g_{\mu\nu}\, ,
$$
where the~$\bar g$-cosmological constant is~$\bar \lambda = \lambda/\gamma$.
The equations of motion~(\ref{eom}) then imply
$
(\Lambda-\lambda) g_{\mu\nu}+\frac{\mu^4}{M^2}\, \tau_{\mu\nu}=0=(\gamma\bar\Lambda-\lambda) \, g_{\mu\nu}
+\frac{\mu^4}{\bar M^2}\, \bar\tau_{\mu\nu}
$.
Since for this configuration,~$\bar\tau_{\mu\nu}=\frac1\gamma \tau_{\mu\nu}=6 g_{\mu\nu}$, we learn
\begin{equation}\label{balance}
M^2(\Lambda - \lambda) + 6\gamma\mu^4=0=\bar M^2(\gamma\bar\Lambda - \lambda)+6\mu^4\, .
\end{equation}
Consequently we must require
$
\lambda=\Lambda+\frac{6\gamma\mu^4}{M^2}
$.
Thus, away from the first of two critical cases (these will coalesce to a single one below), where~$\mu^4\neq\frac{M^2  \bar\Lambda}6$ we find 
$$\gamma=\frac{\Lambda-\frac{6\mu^4}{\bar M^2}}{\bar \Lambda-\frac{6\mu^4}{ M^2}}\, ,$$
while for~$\mu^4=\frac{M^2  \bar\Lambda}6$, the parameter~$\gamma$ remains undetermined. The second critical case at
~$\mu^4=\frac{\bar M^2  \Lambda}6$ appears because it yields an illegal~$\gamma=0$  rescaling (this also follows by 
symmetry in the two metrics).

Having determined the parameter conditions for existence of the ``diagonal'' BMG Einstein solution, we now 
study its fluctuations.
Linearizing 
%about a diagonal bimetric Einstein solution:
$$
g_{\mu\nu}=g^E_{\mu\nu}+h_{\mu\nu}\, ,\qquad
\bar g_{\mu\nu}=\gamma g^E_{\mu\nu}+\bar h_{\mu\nu}\, ,\qquad
$$
we find~$\varphi_{\mu\nu}\approx \sqrt{\gamma}\, g^E_{\mu\nu}+\frac{\sqrt{\gamma}}2  h_{\mu\nu}+\frac12\frac1{\sqrt\gamma} \bar h_{\mu\nu}$
and thus
\begin{equation*}
\begin{aligned}
\tau_{\mu\nu}\ &\approx \  6\gamma g_{\mu\nu}^E  + 8\gamma\, (h_{\mu\nu}-\frac14g^E_{\mu\nu}h )  -2\, (\bar h_{\mu\nu}-g^E_{\mu\nu} \bar h)\, ,\\
\bar \tau_{\mu\nu}\ &\approx\,   \ 6 g_{\mu\nu}^E \ -\, 2\, (h_{\mu\nu}-g^E_{\mu\nu} h)\ +\ \frac{8}\gamma\, (\bar h_{\mu\nu}-\frac14g^E_{\mu\nu}\bar h )\, .
\end{aligned}
\end{equation*}
Here index   manipulations and traces are performed with the Einstein background metric~$g^E_{\mu\nu}$.
Orchestrating the above, we obtain (in a matrix notation) the field equations,
\begin{widetext}
\begin{equation*}
\begin{pmatrix}
G_{\mu\nu}^L(h)-\lambda h_{\mu\nu}\\[3mm]
G_{\mu\nu}^L(\bar h)-\lambda \bar h_{\mu\nu}
\end{pmatrix} \ =\  
\frac{2\mu^4}{M^2\bar M^2}
\begin{pmatrix}
\gamma \bar M^2 & -M^2\\[3mm]
-\gamma \bar M^2& M^2
\end{pmatrix}
\begin{pmatrix}
h_{\mu\nu}-g_{\mu\nu}^E h\\[3mm]
\bar h_{\mu\nu}-g_{\mu\nu}^E \bar h
\end{pmatrix}\, .
\end{equation*}
\end{widetext}
The~$2\times2$ ``mass'' matrix has eigenvalues~$0$ and~$M^2+ \gamma\bar M^2$. Thus, away from the disallowed values of~$\mu$ given above,
we find a (massless, massive) graviton pair. 
%with mass
%$$
%m^2=4\mu^4\, \Big[\frac1{\bar M^2}+\frac{\gamma}{M^2}\Big]
%$$
The linearized graviton mass parameter~$m^2=4\mu^4(\bar M^{-2}+\gamma\ M^{-2})$
can still be changed arbitrarily by making further constant metric
rescalings. For example, sending~$g_{\mu\nu}^E\to\beta g_{\mu\nu}^E$ changes the~$(m^2,\lambda)$ system to~$(\beta m^2, \lambda)$.
[The choice~$\beta=\frac1{\sqrt{\gamma}}$ gives, for example, a mass that is symmetric in barred and unbarred quantities.]
Precisely at the critical cases, this freedom drops out:
Consider the critical point,~$\mu^4=\frac{M^2\bar\Lambda}{6}$. Although~$\gamma$ is undetermined, we are still forced by Eq.~(\ref{balance}) to set
$$
M^2\bar\Lambda = \bar M^2 \Lambda\, ,
$$
which yields equivalence of the two critical~$\mu$ values.
The cosmological constant of the background is now
$$
\lambda=\Lambda+\gamma\bar \Lambda\, .
$$
For the linearized equations of motion we find
\begin{equation*}
\begin{pmatrix}
G_{\mu\nu}^L(h)-\lambda h_{\mu\nu}\\[3mm]
G_{\mu\nu}^L(\bar h)-\lambda \bar h_{\mu\nu}
\end{pmatrix} \ =\  
\frac{1}{3}
\begin{pmatrix}
\gamma \bar \Lambda & -\Lambda\\[3mm]
-\gamma \bar \Lambda&\Lambda
\end{pmatrix}
\begin{pmatrix}
h_{\mu\nu}-g_{\mu\nu}^E h\\[3mm]
\bar h_{\mu\nu}-g_{\mu\nu}^E \bar h
\end{pmatrix}\, .
\end{equation*}
which describes  a massless graviton as well as a mode with mass
$$
m^2=\frac{2\lambda}{3}\, .
$$
This is precisely the partially massless tuning and fuels hope that the full BMG
may enjoy an (interacting) partially massless limit~\cite{HassanPM}. 

Diagonalizing the field equations, we find that~$h^{\rm grav}_{\mu\nu}:=h_{\mu\nu} + \bar h_{\mu\nu}$
describes the linearized, cosmological,  graviton excitation,~$G_{\mu\nu}^L(h^{\rm grav})-\lambda h^{\rm grav}_{\mu\nu}=0$, 
while~$h^{\rm PM}_{\mu\nu}:=\gamma \bar \Lambda h_{\mu\nu}-\Lambda\bar h_{\mu\nu}$ is the PM mode which obeys
$$
{\cal G}^{\rm PM}_{\mu\nu}:=G_{\mu\nu}^L(h^{\rm PM})-\frac{4\lambda}{3}\tilde h^{\rm PM}_{\mu\nu}=0\, ,
$$
(where~$\tilde h_{\mu\nu}:=h_{\mu\nu} - \frac14 g^E_{\mu\nu} h$). In particular,
the linear  Bianchi identity is
$$
\big(\nabla^\mu\nabla^\nu+ \frac \lambda 3 \, g_E^{\mu\nu}\big) {\cal G}^L_{\mu\nu}(h^{\rm PM})\equiv 0\, ,
$$
and corresponds to the PM invariance $\delta h_{\mu\nu}=\big(\nabla_\mu\nabla_\nu+ \frac \lambda 3 \, g^E_{\mu\nu}\big)\alpha(x)$.
As a forewarning, we note that the combination of linearized fluctuations in~$h_{\mu\nu}^{PM}$ depends on the undetermined parameter~$\gamma$, which will limit the usefulness of the linear analysis when searching for a non-linear extension.

\section{No PM BMG?}\label{problems}

Our main   argument against PM BMG is  that a Bianchi identity and associated non-linear gauge invariance (at some
 critical point), extending
the linear one exhibited above, is unlikely to exist.
One problem is that no covariant formulation of the BMG scalar constraint
responsible for the~$7=2+5$ DoF count of an interacting graviton and massive spin~2 excitation has yet been found: to date all groups
claiming that the sixth, massive ghost-like, excitation is absent for distinguished BMG mass terms relied on a 3+1 ADM-type analysis~\cite{Hassan1}. Strangely enough, although these analyses claim to prove that BMG propagates~$7$ DoF, none of them 
are able to show that a PM parameter-tuning yields a~$6$ DoF PM limit, even though for cosmological PF models one sees this quite directly
(in fact, in a~$3+1$ split, the zero helicity part of the action becomes exactly zero at the PM point~\cite{PMstab}).
On the one hand, we suspect that an exhaustive PM BMG no-go proof will
require a computation of the Poisson bracket algebra of the (non-covariant) constraints in order to check for an enhancement of the first class symmetry algebra.  If there really is no covariant formulation of the scalar constraint, however,\footnote{Clever use of auxiliary fields might perhaps permit lifting the constraints to covariant form, but that need not necessarily be useful in itself.}, then it is already very difficult to believe that a covariant 
gauge symmetry arises--recall that in the free theory, the partially massless gauge symmetry corresponds to a double derivative Bianchi identity, obtained from a mass~:~cosmological constant tuning limit of the double divergence constraint\footnote{Here, we are assuming that any putative PM gauge invariance (and thus Bianchi identity) is covariant with respect to diffeomorphisms. One might in principle envision a situation where the commutator of diffeomorphisms and PM gauge symmetries vanishes on-shell, thus allowing a non-covariant PM gauge invariance without introducing new first class symmetries. We strongly doubt that any such non-covariant PM gauge invariance exists, however.  }.

To see why mGR enjoys a covariant system of constraints, while BMG seems not to, consider a 
putative non-linear Bianchi identity of the form~$$\nabla\stackrel{\scriptscriptstyle ?}{.}\nabla \stackrel{\scriptscriptstyle ?}{.} {\cal G} - \bar \nabla\stackrel{\scriptscriptstyle ?}{.}\bar \nabla\stackrel{\scriptscriptstyle ?}{.}\bar {\cal G}\  -\   {\rm l.o.t.}\equiv 0\, ,$$
where the lower order terms (l.o.t.) involve fewer derivatives on the field equations~${\cal G}$ and~$\bar {\cal G}$. On general grounds, this identity can involve no more than two covariant derivatives. We have denoted the contraction of these on the field equations by 
$\stackrel{\scriptscriptstyle ?}{.}$  to indicate that the tensor contraction of indices here is still to be determined. We can study the identical vanishing of this quantity order by order in derivatives on the two dynamical fields~$g_{\mu\nu}$ and~$\bar g_{\mu\nu}$. In particular 
we can first focus on~$g_{\mu\nu}$. This allows us to steal results from the known mGR case.
There, one knows that  the contraction of the first covariant derivative is the one giving the vector constraint/Bianchi identity, namely
$g^{\mu\rho}\nabla_\mu{\cal G}_{\rho\nu}$. In other words, the Bianchi identity should follow from an appropriate divergence~$\nabla\stackrel{\scriptscriptstyle ?}{.} {\cal C}$ of the
vector constraint (recall that~${\cal C}_\mu$ and~$\bar {\cal C}_\mu$ are not independent)
$$
{\cal C}_\mu\ \stackrel{g}=-\, \frac{2\mu^4}{M^4}\,  \varphi^{\nu\rho}
\big(K_{\nu\rho\mu}-g_{\nu\rho} K_{\sigma}{}^\sigma{}_\mu
+g_{\mu\nu}K_{\sigma}{}^\sigma{}_\rho\big)\, .
$$
Again, by working order by order in derivatives on the independent field~$g_{\mu\nu}$ we deduce from the mGR 
case that any putative Bianchi identity follows from the particular choice of divergence~$g^{\mu\nu}\nabla_\mu {\cal C}_\nu$.
This quantity has been computed in~\cite{DSW}, but here we need only focus on terms involving two derivatives on the dynamical
variables which can be computed using the following identity relating Riemann tensors of differing connections to the contorsion:
$$
R_{\mu\nu}{}^m{}_n(\omega)-R_{\mu\nu}{}^m{}_n(\bar \omega)=2\big(\nabla_{[\mu}K_{\nu]}{}^m{}_n-K_{[\mu}{}^m{}_{|r|}
K_{\nu]}{}^r{}_n\big)\, .
$$
Using this, we find at leading derivative order
\begin{widetext}
$$
g^{\mu\nu}\nabla_\mu {\cal C}_\nu\stackrel{g}=-\, \frac{2\mu^4}{M^4}\, 
\Big(\varphi^{\nu\rho}\big[R(\omega)_{\mu\nu\rho}{}^\mu-R(\bar\omega)_{\mu\nu\rho}{}^\mu\big]
+\frac12\, \varphi^\rho_\rho\, \big[R(\omega)_{\mu\nu}{}^{\nu\mu}-R(\bar\omega)_{\mu\nu}{}^{\nu\mu}\big]\Big)+{\rm l.o.t.}\, .
$$
\end{widetext}
There are various lessons to be learnt from this relation: First, for mGR where~$R(\bar\omega)_{\mu\nu}{}^m{}_n$
is the Riemann tensor of the fiducial background, the existence of the scalar constraint is immediate, because
the Riemann tensor for~$g_{\mu\nu}$ appears only in its Ricci or Ricci-scalar form, both of which can be canceled at leading 
derivative order by adding a trace of the field equation~${\cal G}_{\mu\nu}$~\cite{DMZ}. If one further takes the background to be constant curvature, the existence of a PM mGR limit hinges on the lower order terms. These were shown not to cancel in~\cite{DSW}, thus ruling out the PM mGR theory. If we were to require that the PM mGR limit also held for Einstein backgrounds (after all, these are sufficient for consistent linear PM propagation~\cite{PM,PMstab}), the terms~$R(\bar\omega)$ already destroy that model. They are, in fact, also lethal for the prospects of a PM BMG. The main point is simple: 
 the Riemann tensor
of the second metric~$\bar g_{\mu\nu}$ does not appear as a Ricci or Ricci-scalar and thus cannot be removed using the field equation~$\bar{\cal G}_{\mu\nu}$. Note, that this does not contradict previous claims that the sixth massive ghost mode is absent, because
the leading double time derivative terms  can be removed by  choosing a time slicing adapted to the metric~$g_{\mu\nu}$\footnote{We are not claiming that the constraint analysis relies on this choice, only that obtaining the scalar constraint from the above combination does.}.

We have experimented with other possible contractions~$\nabla\stackrel{\scriptscriptstyle ?}{.} {\cal C}$, as well as using identities
for the contorsion following from the relation~(\ref{symmetry}) and found no way to avoid the appearance of ``bare'' Riemann tensors
in the putative Bianchi identity. These cannot be converted to Ricci and Ricci-scalar tensors nor in turn traded for equations of motion. 
By way of warning, even supposing that a combination of double divergences and traces of field equations canceling all second derivatives of 
$g$ and~$\bar g$ existed, a Bianchi {\it identity} further requires all remaining terms to cancel {\it off shell}. This failed spectacularly for the mGR model. Moreover,  the coupling of two metrics produces a new covariant tensor--the contorsion, and there is no mechanism (aside from conformal symmetry which yields relatively ghost CG) preventing this
from happening again. [However, unlike our previous study of PM mGR~\cite{DSW}, we have not conducted an order by order expansion in fields and derivatives.]

Aside from the linear analysis of~\cite{HassanCG} that we summarized in Section~\ref{linear}, the other piece of purported PM BMG 
evidence is that a certain truncation of a higher derivative ``version'' of (tuned) BMG yields CG. By virtue of its local conformal and diffeomorphism invariances, CG (non-linearly) propagates  PM and graviton, relative-ghost, modes~\cite{Maldacena,Joung}.
However, the higher derivative model studied in~\cite{HassanCG} is {\it not} equivalent to 
 BMG because it is obtained by the generally illegitimate 
 procedure of plugging dynamical field equations back into the action principle\footnote{This point is actually conceded in an Appendix of~\cite{HassanCG}, but there it is suggested---based on the example of a Gaussian path integral of a two-scalar, free theory---that by carefully keeping track of sources, ghosts could be somehow removed. In any case, it is inconsistent to perform interacting path integrals by going (even partially) on-shell, that is  by replacing fields by  solutions to their field equations.}: Consider  a simple two-field model ~$S[\phi,\psi]=\int\big(\frac12\phi\square \phi + \frac 12\psi \square \psi - \phi \psi\big)$ (in obvious mass units);
the equations of motion imply~$\square \phi =\psi$ and~$\square \psi=\phi$. Thus one might be tempted to conclude---by using~$\psi=\psi(\phi):=\square \phi$---that an equivalent model is~$S[\phi]=S[\phi,\psi(\phi)]=-\int\frac12\phi[\square-\square^3] \phi$ with equations of motion~$[\square-1][\square+1]\square \phi=0$. However this latter model has a solution~$\square \phi=0$ that does not solve the original system of equations.
That the same inconsistency afflicts the higher derivative model obtained from the bimetric theory is not hard to see.
For the bimetric theory this ``half-shell'' procedure yields CG with higher order curvature  corrections:~$S=\int W^2 + {\cal O}(R^3)$ (where $W$ denotes the Weyl tensor).
So any conformally Einstein manifold is a leading order solution,  in particular~$G_{\mu\nu}(g)=\Lambda g_{\mu\nu}$. The bimetric equations of motion have the form~$G_{\mu\nu}(g)-\Lambda g_{\mu\nu}=\frac{\mu^4}{M^2}\, \tau_{\mu\nu}(g,\bar g)$ plus the same with bars. Clearly we are now forced into the same corner, namely~$\bar g_{\mu\nu}=0$ and in turn\footnote{
  In fact, PM invariance of the higher derivative model is intimately tied  to the PM mGR Bianchi identity, as can be seen by  applying
 the 1.5 order formalism (see~\cite{Supergravity}):
The higher derivative model is obtained by solving for one metric~$\bar g$ (say) in terms of the other by its algebraic
field equation:
$
\frac{\delta S(g,\bar g)}{\delta g} = 0 \: \Rightarrow \: \bar g=\bar g(g)$.
Plugging this relation back in the action gives a higher derivative model~$S(g):=S(g,\bar g(g))$ (whose leading
term is claimed to be the CG action~\cite{HassanCG}) can then easily be varied 
$$
\delta S(g) = \frac{\delta S(g,\bar g(g))}{\delta g} \delta g + \frac{\delta S(g,\bar g)}{\delta \bar g}\Big|_{\bar g=\bar g(g)} \frac{\delta \bar g(g)}{\delta g}\delta g $$
$$
\hspace{-1.3cm}= \frac{\delta S(g,\bar g)}{\delta \bar g} \Big|_{\bar g=\bar g(g)}\frac{\delta \bar g(g)}
{\delta g}\delta g\, . 
$$
The~$g$ variation  vanishes thanks to the 1.5 order method (because~$\bar g$ obeys the field equations~$\frac{\delta S(g,\bar g)}{\delta g}=0$).
Moreover,~$\frac{\delta S(g,\bar g)}{\delta \bar g}=0$ is exactly the equation of motion of mGR  for~$\bar g$. 
The functional relation~$\frac{\delta S(g,\bar g)}{\delta \bar g} \delta \bar g=0$ (for general~$\bar g$) amounts precisely to a PM mGR Bianchi identity. 
Its absence  has by now been independently established by three groups~\cite{Zinoviev,DSW,deRhamPM}
(although conceivably,  the PM mGR Bianchi identity failure could vanish on the  subspace~$\bar g= \bar g (g)$).}~$g_{\mu\nu}=0$. The fact that a certain truncation of a higher derivative theory---that itself is explicitly INequivalent to BMG---has a scalar gauge invariance
does not imply a scalar gauge invariance of PM BMG.

Finally, even supposing we were to employ the  higher derivative model as inspiration for a possible PM BMG gauge principle,  since its action starts with that of CG, a putative
PM gauge transformation would then begin with a Weyl transformation. Higher  derivative corrections to these transformations can then be ignored
by taking the special case where the gauge parameter is constant. However, this would imply a rescaling symmetry, which can hardly be the case when higher order terms involve inverse powers of a dimensional constant--the cosmological constant.
(One might hope to modify this scaling symmetry by some sort of  curvature-dependent extension reducing to unity in 
a low curvature limit, but this seems a pointless exercise given our previous no-go arguments.) We therefore conclude, with some confidence, that GR is indeed an isolated theory, like its non-abelian YM cousin.

\section{Conclusions}

We have argued that the existence of a PM version of BMG is crucial to its consistency.
In particular a consistent PM could shield the model from causal
inconsistencies of the type uncovered for mGR in~\cite{DW,DSW}. 
We have further argued that the known no-go results for PM mGR make survival of a PM symmetry in BMG most unlikely. Supposing, as we believe, that this can be made rigorous, what are the prospects for improving BMG?
Perhaps some string-inspired approach such as that of~\cite{Porrati} (which attempts to restore consistency of  massive charged $s=3/2,2$ interactions by embedding them in a consistent underlying framework) may help. (First steps in this direction have been made in~\cite{Kiritsis}, although even there
instabilities are encountered.) This would likely 
entail also modifying the kinetic, Einstein curvature, term (as recently attempted in~\cite{HB1}).
Maintaining even the BMG vector constraint is difficult in such an approach, so the obstacles here seem high\footnote{Since our work's posting, absence of a PM version of BMG was independently reported in~\cite{Fasiello}}.

\begin{acknowledgments}
A.W. thanks G. Gabadadze, F. Hassan, K. Hinterbichler and  M.S and A.W. thank  N. Kaloper for discussions.  S.D. was supported in part by NSF PHY- 1266107 and DOE DE- FG02-164 92ER40701 grants.
\end{acknowledgments}

\end{document}